\documentclass[prl, twocolumn, showpacs, byrevtex]{revtex4}
\input epsf
\usepackage{graphicx}
\usepackage{dcolumn}
\usepackage{bm}
\begin{document}
\preprint{}
\title{BASIC PROBLEMS OF A MICROSCOPIC THEORY OF A MANY BODY
QUANTUM SYSTEM}
\author{Y. S. Jain}
\email{ysjain@nehu.ac.in}
\email{ysjain@email.com}
\affiliation{Department of Physics, N.E.H.U., Shillong-793 022, Meghalaya,
India}
\date{today}
\begin{abstract}
Basic problems of a microscopic theory of many body quantum systems
and different aspects of a new approach which can help in solving them
are discussed in detail.   To this effect we make a critical study of
the wave mechanics of two hard core quantum particles and discover its
several untouched aspects, \textit{viz.} : (i) the useful details of
$\psi_k(r)$ (representing the relative motion of two particles), (ii) the
expectation value of \textit{hard core} (HC) repulsion ($<V_{HC}(r)>$),
(iii) the inconsistency of the statements, $r \le \sigma$
and $\psi_k(r \le \sigma) = 0$ ($\sigma =$ HC diameter of a particle),
with uncertainty principle particularly for low $k$ values, (iv) the lower
bound of allowed values of $k = 2q$, (v) the dominance of interparticle
phase correlation in low temperature phase.  For the first time
this study concludes that $<V_{HC}(r)>$ has \textit{zero} value which
does not agree with its \textit{non-zero} value known for the last several
decades.  This also finds compelling reasons for a system of interacting
bosons such as liquid $^4He$ to have (\textbf{q}, -\textbf{q}) pair
condensation with allowed $q$, obviously controlled by $V_{HC}(r)$,
to satisfy $q \ge \pi/d$.  Several important aspects of $N$ body quantum
systems like liquids $^4He$ and $^3He$ are also concluded.  Free from any
error [see editor's note \textit{J. Scientific Exploration} \textbf{ 16}(1), p.1
(2002)], our approach can help in developing nearly exact microscopic
theories of widely different systems of interacting bosons and fermions, as
demonstrated for liquids $^4He$ type systems [\textit{J. Scientific
Exploration}, \textbf{ 16}, 77-116 (2002)].  The paper also sums up the
expert observations with our response to facilitate one to have a
critical assessment and better understanding of the new approach. 
\end{abstract}
\pacs{67.40.-w 67.57.-z}
\maketitle
\section{1.0 Introduction}
Microscopic studies of a \textit{system of interacting bosons} (SIB) such
as liquid $^4He$ (LHe-4) and a \textit{system of interacting fermions} (SIF)
such as liquid $^3He$ (LHe-3) are of great fundamental importance \cite{1,2}
because these systems provide an opportunity to investigate
quantum effects at macroscopic level.  The scope of such studies has
been widened by the recent experimental discovery of Bose Einstein
condensate (BEC) in trapped dilute gases because interparticle
interactions in these gases too play as important role
as in LHe-4 \cite{3}.  While the basic aspects of the subject are elegantly
discussed in several texts \cite{4,5,6,7}, the wealth of experimental and
theoretical results generated from widely different studies of various
SIB and SIF has been reviewed recently in several reports \cite{8,9},
books \cite{10, 11} and research articles \cite{12, 13}.  References to other
important publications and review articles of older times can be traced from
all these sources and our recent publications \cite{14,15,16} presenting
a new approach to the microscopic understanding of a SIB and its
unification with that of a SIF.
\par
It appears that theories of these systems, developed by using different
mathematical tools and varying sets of assumptions and conjectures, can be
placed into two groups, \textit{viz.}, G1 group of the conventional theories
which assume the existence of $p=0$ condensate, $n_{p=0}(T)$ (a macroscopic
fraction of particles condensed into a single particle state of $p=0$), as
the basic origin of superfluidity of a SIB \cite{8,10,11} and the
formation of a kind of Cooper pairs as the main factor for the superfluidity
of a SIF \cite{9,10}, and G2 group of a recently developed theoretical
framework which uses a new approach to formulate the microscopic theories
of a SIB \cite{15} and SIF \cite{17}.  The approach makes no assumption
like the existence of $p=0$
condensate for a SIB or the formation of Cooper like pairs for a SIF.  In
stead it uses the results that we obtain by solving the $N$ body
Schr\"{o}dinger equation.  It concludes that (\textbf{ q}, -\textbf{ q}) bound
pairs form an important basis for the unique properties
of superfluid phase of a SIB/SIF \cite{15,16,17}.
\par
In one of the two approaches of conventional theories, one starts with
the Hamiltonian of the system written in terms of second quantized
Schr\"{o}dinger field and then proceeds to solve the problem by using
several important inferences such as : (i) the HC potential can be used
perturbatively by pseudo-potential method, (ii) coupling constant can be
related directly to the two body scattering length ($a$), (iii) a
dimensionless expansion parameter ($n.a^3$) can be used to perform
perturbation calculation, \textit{etc.} \cite{1,2,3,4,5, 8}.
However, in an alternative approach of variational calculations based on
Jastrow correlation or Feenberg approximation one finds the radial
distribution function, $g(r)$, and liquid structure factor, $S(Q)$, which
can be used to calculate the \textit{ground state} (G-state) properties,
excitation spectrum, different thermodynamic properties and equation
of state \cite{12,13}.  Widely different methods of computer calculation of
$g(r)$, $S(Q)$ and other properties are discussed and reviewed in \cite{18}.
In spite of highly complex mathematical formulations and calculations, 
these approaches are believed to have succeeded in obtaining certain
results in support of their presumptions of the existence of $p=0$
condensate in superfluid $^4He$ and formation of (\textbf{q}, -\textbf{q}) bound
pairs in case of $^3He$ as a basis of superfluidity of these systems.
However, several experts in the field (\textit{e.g.} Rickayzen \cite{19},
Woods and Cowley \cite{20} and Sokol \cite{21} in case of $^4He$ and
Senatore and March \cite{9} in case of $^3He$) also underline the
limitations and difficulties of these approaches in providing theories that
explain the experimental properties of a SIB/SIF.
\par
Our new approach, on the other hand, is a simple approach.
It follows the standard method of solving the Schr\"{o}dinger equation
of the system and uses the wave mechanics of two HC particles as its
basis.  We make a critical study \cite{14} of this mechanics to discover its
several untouched aspects which serve as the basic foundations of our
approach.  We find \cite{15,16,17} that : (i) each particle in a SIB/SIF
represents (\textbf{q}, -\textbf{q}) pair moving with \textit{center of
mass} (CM) momentum \textbf{K}, (ii) with the onset of superfluid transition
these particles form (\textbf{q}, -\textbf{q}) bound pairs with their
phase positions locked at a relative separation of $\delta \phi = 2n\pi$; it
happens with every particle representing a (\textbf{q}, -\textbf{q}) pair
and leads the system to have a kind of collective binding and a
macro-molecular behavior, (iii) while the $\lambda$-transition of a SIB
represents the condensation of particles as (\textbf{q}, -\textbf{q}) pairs
in the G-state of the system defined by $q = \pi/d$ and $K = 0$ \cite{15},
the superfluid transition of a SIF \cite{16,17} represents the fall of 
(\textbf{q}, -\textbf{q}) pairs in a state of $q = \pi/d$ and $K$ ranging
between $K = 0$ and $K = K_F$ (with $K_F$ being the Fermi wave vector);
the formation of (\textbf{q}, -\textbf{q}) bound pairs in both cases
arises as a combined effect of interparticle phase correlations and
interparticle attraction \cite{15,16,17}.
\par
As demonstrated in \cite{15}, the new approach provides an almost exact
theory of a SIB that explains the properties of liquid $^4He$ with unparalleled
accuracy, simplicity and clarity.  It has several advantages.  It provides
a framework that can unify the physics
of widely different SIB (\textit{e.g.} low dimensional systems, BEC state
of dilute gases, \textit{etc.} and SIF (\textit{e.g.} superfluid $^3He$,
atomic nucleus, supercondutors, \textit{ etc.}) \cite{16,17}.  It does not
require complex mathematical formulations and involved computer
calculations.  Evidently, these results provide enough reasons
to search for the origin of the difficulties with our conventional
approaches so that a comprehensive approach free from every problem
can be identified for future studies.  To facilitate
the process we make a critical study of certain physical realities of
LHe-4 anf LHe-3 systems and intricacies of wave particle duality and
conclude that certain aspects of conventional approaches are in error
and the new approach can help in developing an almost exact
microscopic theory of these systems.  
\par
The paper has been arranged as follows.  While Section 2.0  defines the
basic nature of a SIB/SIF to which our approach can be applied, wave
mechanics of two HC particles, which serves as an important
basis of a theory of a SIB/SIF, is critically analyzed in Section 3.0.
The way this analysis modifies our conventional understanding of a
$N$ body system is discussed in Section 4.0 and a summary of this study
is presented in Section 5.0.  While the paper is basically an expanded
version of \cite{14}, it also analyzes some important aspects of $N$ body
systems for the first time.  For example, by using a systematic mathematical
derivation, this paper finds (\textit{cf.} Section 4.5) that the expectation
value of $V_{HC}(r_{ij})$ for every state of $N$ body SIB/SIF has zero value.
This has a direct conflict with non-zero value of $<V_{HC}(r_{ij})>$
\cite{4,5}, that we know for the last seven decades.  Similarly the
allowed $q$ values satisfy $q \ge \pi/d$ indicating that no particle in
a SIB/SIF has $q = 0$ as presumed in conventional formulations.
Unfortunately, in spite of the fact that our approach provides a theory
that explains experimentally observed properties of superfluid $^4He$
(even those which find no explanation in conventional theories), the
accuracy and importance of our work has been ignored.  In stead it faced
unexpected difficulties with experts \textit{not for any error} \cite{22} but
for the conflict of its basic ideas/results with those of their conventional
framework.  The history of the development of science witnesses many
instances when non-conventional ideas/suggestions
faced such difficulties before their acceptability.  Consequently,
we sum up all these difficulties and our response \cite{23} so that the
accuracy
of different aspects of our work could be critically assessed and better
understood.  We hope that our approach should provide scientifically
sound answers to different possible questions.  It is satisfying that
several competent people found merit in our work \cite{22}.
\section{2.0  Basic Nature of Our SIB/SIF}
To a good approximation, interparticle interaction in LHe-4/LHe-3
type system can be represented by two-body interaction $V(r_{ij})$ by
presuming that the spin-spin type interactions (if exist) can be treated
separately as a perturbation after solving the equation
\begin{equation}
\left[-{\hbar^2\over{2m}}\sum_i^N
{\bigtriangledown}^2_i + \sum_{i<j} V(r_{ij})\right]{\Psi}_n =
E_n{\Psi}_n \label{eq1}
\end{equation}
\noindent
for its allowed states.  $V(r_{ij})$ can be considered as the sum of
: (i) the repulsive potential $V^{\rm R}(r_{ij})$ approximated to
a HC interaction $V_{HC}(r_{ij})$ defined by $V_{HC}(r_{ij} \ge \sigma)
= 0$ and $V_{HC}(r_{ij} < \sigma) = \infty$ with $\sigma$ being the
HC diameter of a particle and (ii) a relatively long range weak attraction
$V^{\rm A}(r_{ij})$.  Since $V^{\rm A}(r_{ij})$ can be replaced by
constant negative potential (say, $-V_o$), atoms in these systems
move freely like HC particles on a surface of potential $-V_o$.  This
not only agrees with the translation symmetry of a liquid \cite{24}, but also
concludes that in formulating the theory of such a SIB/SIF we, at the first
stage, have to deal with $V_{HC}(r_{ij})$ and $V^{\rm A}(r_{ij})$ can be
used as a perturbation at the second stage.  As such
the effective $N$ body Hamiltonian, with which a first stage analysis
should be started, can be expressed by
\begin{equation}
H^o(N) = -{\hbar^2\over{2m}}\sum_i^N
{\bigtriangledown}^2_i + \sum_{i<j} V_{HC}(r_{ij}). \label{eq2}
\end{equation}
\noindent
However, following a discussion in Section 4.8, it is evident that the
effective interaction incorporated in our new approach hardly differs
from the real interaction between the particles.
\section{3.0  Wave Mechanics of Two HC particle}  
\par
\noindent
\textit{3.1. Schr\"{o}dinger equation} : In what follows from the above
discussion, the wave mechanics of two particles (say P1 and P2) in a
above defined SIB/SIF can be described by
\begin{equation}
\left[-{\hbar^2\over{2m}}
\left({\bigtriangledown}^2_1
+ {\bigtriangledown}^2_2\right) + V_{HC}(r)\right]{\Psi}(r_1, r_2) =
E{\Psi}(r_1, r_2). \label{eq3}
\end{equation}
\noindent
For a central force like $V_{HC}(r)$, Eqn.(3) can also be
expressed \cite{25} as
\begin{equation}
\left[-{\hbar^2\over{4m}}{\bigtriangledown}^2_R 
-{\hbar^2\over{m}}{\bigtriangledown}^2_r + V_{HC}(r)\right]{\Psi}(r,R) =
E{\Psi}(r,R) \label{eq4}
\end{equation}
\noindent
with 
\begin{equation}
{\Psi}(r,R) = {\psi}_k(r)\exp{[i(\textbf{K}.\textbf{R})]} \label{eq5}
\end{equation}
\noindent
describing a state of P1 and P2 having relative momentum $\textbf{k}$
and CM momentum $\textbf{K}$.  Here ${\psi}_k(r)$
representing the relative motion of P1 and P2 satisfies
\begin{equation}
[-({\hbar}^2/{m}){\bigtriangledown}^2_r + V_{HC}(r)]{\psi}_k(r) =
E_k{\psi}_k(r) \label{eq6}
\end{equation}
\noindent
with $E_k = E - {\hbar}^2K^2/4m$.  Different notations in Eqns.(3-6)
including 
\begin{equation}
\textbf{r} = \textbf{r}_2 - \textbf{r}_1 \quad {\rm and} \quad
\textbf{k} = \textbf{k}_2 - \textbf{k}_1 \label{eq7}
\end{equation}
\begin{equation}
\textbf{R} = (\textbf{r}_1 + \textbf{r}_2)/2 \quad {\rm and} \quad
\textbf{K} = \textbf{ k}_1 + \textbf{k}_2 \label{eq8}
\end{equation}
\noindent
have their usual meaning.  It is evident that while allowed \textbf{k} can
be affected by $V_{HC}(r)$, similar values of \textbf{K} representing a
kind of free particle motion ($\exp{[i.(\textbf{K}.\textbf{R})]}$) are
controlled only by the dimensions
of the container.  This difference in the nature of \textbf{k} and \textbf{K}
motions of a pair clearly indicates that these motions can get totally
delinked under certain physical conditions of a SIB/SIF.  In fact, as
found in our recent study of a SIB \cite{15}, this really happens below
$\lambda$-point and its superfluid phase is concluded to behave like a
homogeneous mixture of two fluids as envisaged exactly by Landau \cite{26}.
\par
\noindent
\textit{3.2.  Characteristic details of ${\psi}_k(r)$ }:
The details of a ${\psi}_k(r)$, as defined by Eqns.(6 and 7), have
so far been considered sufficient for most theoretical formulations of a
SIB/SIF.  However, we find no reason to ignore its additional details
which not only simplify a theoretical formulation but also provide a
better understanding of the system.  To this effect we analyze
${\psi}_k(r)$ as seen from a frame attached to their CM defined by
\textbf{R} = 0 and \textbf{K} = 0 and find that 
\begin{equation}
\textbf{k}_1 = -\textbf{k}_2  = \textbf{q} \label{eq9}
\end{equation}
\noindent
and
\begin{equation}
\textbf{r}_{CM}(1) = -\textbf{r}_{CM}(2) \label{eq10}
\end{equation}
\noindent
define the characteristic details of such a ${\psi}_k(r)$.  Here
$\textbf{ r}_{CM}$ and $\textbf{ q}$, respectively, represent the position
and momentum of a particle with respect to the CM of the pair.  It is
evident that P1 and P2 in their relative dynamics have: (i) equal
and opposite momenta (\textbf{q}, -\textbf{q}) (\textit{cf.} Eqn.(9)) and (ii)
they maintain a center of symmetry at their CM (\textit{cf.} Eqn.(10)).
In other words, two particles in a laboratory frame could be
identified as a pair having (\textbf{q}, -\textbf{q}) momenta at their CM
which by itself moves with momentum \textbf{K}.  
\par
\noindent
\textit{3.3.  Equivalence of $V_{HC}(r)$ and $\delta{(r)}-$ repulsion }:
Note that two particles of momenta $\textbf{k}_1$ and $\textbf{k}_2$ collide with
each other at their CM defined by $r = 0$ and after this collision they
either bounce back with an exchange of $\textbf{k}_1$ and $\textbf{k}_2$ or they
appear to exchange their positions.  This aspect of the dynamics of P1 and
P2 does not change even if $\sigma$ had been infinitely small (\textit{i.e.}
$\delta-$size).  Only the probability of their collision should change with
a change in $\sigma$ and it would be higher for higher $\sigma$.  Evidently,
two colliding particles can be identified to behave effectively as particles
of $\delta$-size hard core or to interact through $\delta{(r)}$-repulsion.
This implies $V_{HC}(r) \equiv A.\delta{(r)}$ where $A$ representing the
strength of $\delta{(r)}$-repulsion is given by \cite{15}
$A = h^2/8md^2$.  Note that this equivalence does not differ significantly
from $V_{HC}(r) \equiv A'.\delta{(r)}$ (with $A' = h^2.a/(\pi m.d^3)
= (8a/\pi{d}).A$ and $a =$ s-wave scattering length) obtained
analytically \cite{4,5}.
\par
\noindent
\textit{ 3.4. Correct form of ${\Psi}(r, R)$ }:  In view of $V_{HC}(r) \equiv
A.\delta{(r)}$, we find that two particles at $r \not = 0$ experience no
interaction because $A.\delta{(r)}$ has zero value at
all these points.  Consequently, these particles can be represented
by plane waves, $u_{k_i}(r_i) = (1/\sqrt{V}).\exp{(i\textbf{k}_i.\textbf{r}_i)}
\exp{[-iE_it/\hbar]}$ and $\psi{(r, R)}$, at all $r \not = 0$, can be
expressed either as
\begin{equation}
\Psi{(r_1, r_2)} = u_{k_1}(r_1)u_{k_2}(r_2) \label{eq11}
\end{equation}
\noindent
when P1 and P2 having $r >> \lambda (\approx 2\pi/q)$ do not interfere
with each other or as
\begin{equation}
\psi{(r_1, r_2)}^{\pm}
= {1\over\sqrt{2}}\left[u_{k_1}(r_1)u_{k_2}(r_2) \pm
u_{k_2}(r_1)u_{k_1}(r_2) \right] \label{eq12}
\end{equation}
\noindent
when they have $r \approx \lambda$.  Eqn.(12) can be arranged to read as 
\begin{equation}
\Psi{(r, R)}^{\pm} = \psi_k(r)^{\pm}
\exp{(i\textbf{K}.\textbf{R})}\exp{[-i(E_k+E_K)t/\hbar]}  \label{eq13}
\end{equation}
\noindent
with
\begin{equation}
\psi_k(r)^+ = \sqrt{2\over{V}}\cos{(\textbf{k}\textbf{r}/2)}.
\exp{[-iE_kt/\hbar]}\label{eq14}
\end{equation}
\noindent
and
\begin{equation}
\psi_k(r)^- = \sqrt{2\over{V}}\sin{(\textbf{k}.
\textbf{r}/2)}\exp{[-iE_kt/\hbar]}.  \label{eq15}
\end{equation}
\noindent
We find that $\psi_k(r)^+$ or $\psi_k(r)^-$ defines a kind of 
\textit{stationary
matter wave} (SMW) which modulates the relative phase position ($\phi =
\textbf{k}.\textbf{r}$) of two particles in $\phi$-space.  As shown in
\cite{15},
this modulation is the basic reason for the interparticle phase coherence 
experimentally observed in superfluid SIB and SIF; note that in optical
lasers too we have cavity modes in the form of standing electromagnetic
waves which modulate the phase positions of photons.  Note that
$\psi_k(r)^+$ and $\psi_k(r)^-$ differ only in terms of the origin
of the $\phi = \textbf{k}.\textbf{r}$ scale by $\pi$.  While $\phi = 0$ for
$\psi_k(r)^+$ represents the center of an anti-nodal region of its
SMW form, the same for $\psi_k(r)^-$ represents a nodal point.  This
concludes $\psi_k(r)^+ \equiv \psi_k(r)^-$.  To understand this
equivalence we note that in the wave mechanical superposition there is no
way to find whether two particles after their collision have bounced back
on their respective sides of their CM or they exchanged
their positions across this point.  Since the former case represents
the self superposition of each particle, it can best be described by
$\psi_k(r)^+$ because $V_{HC}(r)$ would not operate in such superposition.
However, the latter case implying mutual superposition of two particles
should be represented $\psi_k(r)^-$ because the corresponding wave function
has to vanish at $r=0$ due to $V_{HC}(r)$ operating between the two 
particles.  One can also use $\psi_k(r)^-$ to represent the self
superposition state of a particle with a understanding that the quantum
spread of the particle starts from $r = 0$ (the nodal point of
$\psi_k(r)^-$ identified with the CM of the pair) to $r = \lambda/2$
on the line joining the locations of P1 and P2.  Evidently, $\Psi{(r, R)}^-$
is the best choice of the waveform that can correctly represent a pair
of HC particles or \textit{a particle as the part of this pair}.
\par
\noindent
\textit{3.5. Correct value of $<V_{HC}(r)>$ }: 
In view of what has been concluded in the above section, we have to
use $\psi_k(r)^-$ (or $\Psi{(r, R)}^-$) to evaluate $<V_{HC}(r)>$
in a state of two HC particles (bosons and fermions alike).  
The anti-symmetry of $\psi{(r, R)}^-$ for the exchange
of two particles, agrees with the observation of Woo \cite{24} that HC
bosons in configuration space behave the way fermions behave in momentum
space.  However, this does not mean that larger systems of HC bosons
and HC fermions have no difference due to intrinsic spin values of
bosons and fermions.  They would differ in terms of the occupancy
of the states of allowed $K$ values.  While any number of boson
pairs can have same $K$, no two pairs of fermions have this liberty ;
note that by using the relations, $\textbf{k}_1 = \textbf{K}/2 + \textbf{q}$, 
$\textbf{k}_2 = \textbf{K}/2 - \textbf{q}$, $\textbf{k}'_1 = \textbf{K}'/2 + \textbf{q}'$
and $\textbf{k}'_2 = \textbf{K}'/2 - \textbf{q}'$, one may find that
$\textbf{k}_1 \not = \textbf{k}_2 \not = \textbf{k}'_1 \not = \textbf{k}'_2$
applicable to the set of four identical fermions is equivalent to
$\textbf{K} \not = \textbf{K}'$ when $\textbf{q} = \textbf{q}'$  and
$\textbf{q} \not = \textbf{q}'$ when $\textbf{K} = \textbf{K}'$.  As such
by using $\psi{(r, R)}^-$ (Eqn.13) as a state function of two HC particles
we find that 
\begin{equation}
<\psi{(r, R)}^-|A.\delta{(r)}|\psi{(r, R)}^-> =
|\psi_k{(r)}^-|^2_{r=0} = 0
\label{eq16}
\end{equation}
\noindent
which differs from its non-zero value reported in \cite{4,5}.  It is evident
that only one of the two values (zero or non-zero) of $<A.\delta{(r)}>$
can be true and one can easily find that Eqn.(16) is correct. 
\section{4.0. Some Basic Results for N Body Systems}
\par
\noindent
\textit{4.1. Validity of Eqn.(3/4) for two} HC \textit{particles in a}
SIB/SIF : Since Eqn.(3/4) basically describes two HC particles in free
space, its validity for two such particles in a SIB/SIF must be established
if we wish to use it as a basis to describe such a SIB/SIF.  To this effect
we note that two particles in a SIB/SIF encounter $V_{HC}(r)$ only when
they collide with each other.  While this elastic collision leads to an
exchange of momenta $\textbf{k}_1$ and $\textbf{k}_2$, the fact remains
that P1 and P2 (before and after their collision) have free particle
motion.  Analyzing another possible situation in which mutually colliding
P1 and P2 also collide with other particle(s), we find that $\textbf{k}_1$
and $\textbf{k}_2$ (or $\textbf{k}$ and $\textbf{K}$) after such
collision may assume new values, $\textbf{k}_1'$ and $\textbf{k}_2'$
(or $\textbf{k}'$ and $\textbf{K}'$) but once again P1 and P2
retain their free particle motion.  As such the interparticle interactions
make a pair embedded in a SIB/SIF scatter/jump from its one state to another
state of possible $\textbf{k}$ and $\textbf{K}$, while such a state of
the pair in free space simply remains unchanged.  Evidently, the basic
nature of the dynamics of a pair in two situations does not differ which
means that the states of P1 and P2 in such a system, in spite of their
interaction with other particles, can be described by Eqn.(3/4).
\par
\noindent
\textit{4.2. Correct boundary condition} : We find that the boundary
condition, $\psi_k{(r \le \sigma)} = 0$ or its equivalent, used
to obtain a $\psi_k{(r)}$ either as a solution of Eqn.(6) or by way
of a choice to construct a $\Psi{(r, R)}$, implicitly presumes
that $r = \sigma$ can be determined precisely (\textit{i.e.} with an
uncertainty $\Delta{r} = 0$) which implies that momentum uncertainty
$\Delta{k}$ is infinitely large.  However, since $\Delta{k} = \sqrt{(<k^2>
- <k>^2)}$ for the pair in $\psi_k{(r)}^-$ (Eqn.15) state can be at
the most equal to $k$ \cite{27}, $\Delta{r} = 0$ would be inconsistent
with uncertainty principle for finite $k$ and we have no pair of particles
in a SIB/SIF having $k = \infty$.  The degree of this inconsistency
assumes prominence, particularly, for particles of low momentum,
\textit{viz.}, $q < \pi/\sigma$ (\textit{i.e.} $\lambda/2 > \sigma$) for which
uncertainty in the positions of each
particle, $\Delta{r} = \lambda/2$, becomes much larger than
$r$.  Evidently, one needs to find a right alternative of
$\psi_k{(r \le \sigma)} = 0$ condition.  In this context we take cognizance
of the fact that a particle in wave mechanics manifests itself as a
\textit{wave packet}(WP) of size ${\lambda}/2$ (\textit{i.e.} a \textit{sphere} of
diameter $\lambda/2$) and because two HC particles do not overlap, their
representative WPs should also have no overlap.  Thus, the separation
($<r>$) between two particles should satisfy $<r> \ge {\lambda}/2$
condition or $k<r> \ge 2\pi$) \cite{27} which also follows from the 
uncertainty relation, $\Delta{k}\Delta{r} \ge 2\pi$, because we expect
$k \ge \Delta{k}$, and $<r> \ge \Delta{r}$.
\par
\noindent
\textit{4.3. Allowed values of $q$} : Since two particles in a LHe-4/LHe-3 
have shortest $<r> = d$ (the average nearest neighbor separation),
$<r> \ge {\lambda}/2$ could be read as $d \ge {\lambda}/2$ which implies
that lower bound of $q$ should be $q = q_o = \pi/d$.  This  does not
support the assumption that particles in a LHe-4/LHe-3 can have $q = 0$ and
there can exist a $q = 0$ condensate $n_{p=0}(T)$ in LHe-4 below
$T_{\lambda}$.  This inference agrees with the fact that the existence of
$n_{p=0}(T)$ in superfluid $^4He$ has not been proved experimentally beyond
doubt \cite{21}.  In fact, as discussed above, this assumption contradicts the
uncertainty relation. The lowest possible $q = q_o = \pi/d$ and
corresponding energy $\varepsilon_o = h^2/8md^2$, so inferred, imply that
each particle in the system exclusively occupies a cavity of volume $d^3$
and agrees with excluded volume condition \cite{28} supposed to be satisfied
by HC particles.
\par
\noindent
\textit{4.4. Interparticle phase correlations} : When a particle manifests
itself as a wave, its phase position $\phi$ becomes more relevant than
its $r$.  Naturally, in the low temperature phase of a SIB/SIF where wave
nature dominates the particle nature, the interparticle phase correlation
defined by $g({\phi}) = |\Psi{(r,R)}^-|^2  = |\psi_k{(r)}^-|^2$ (an
explicit consequence of wave nature) should be more relevant than
interparticle position correlation defined by $g(r)= |\phi_k{(r)}|^2$
(a consequence of interparticle potential) where $\phi_k{(r)}$ is the
solution of hyper-netted chain Schr{\"o}dinger equation \cite{12, 13}.  The
fact that this inference also applies to a LHe-3 is well evident since the
details of $\psi_k{(r)}^-$ describing relative dynamics of two
particles does not differ for a pair of HC bosons and that of fermions.
$g({\phi})$ can also be expressed by an equivalent quantum correlation
potential \cite{15} $U = -k_BT_o.\ln|\psi_k{(r)}^-|^2 =
-k_BT_o.\ln{2\sin^2(\textbf{k}.\textbf{r}/2)}$ (with $T_o$ being the
temperature equivalent of $\varepsilon_o$) having minimum
($= -k_BT_o.\ln{2}$) and maximum (=$\infty$) periodically located at
$\phi = \textbf{k}.\textbf{r} = (2n+1)\pi$ (midpoint of an anti-nodal
region a SMW) and $2n\pi$ (location of a node of SMW), respectively.  As
such this potential locks the particles at phase position separation
$\Delta\phi = 2n\pi$ (with $n =$
1, 2, 3, ...) and explains the origin of the experimentally observed
phase coherence in the motion of particles in a superfluid phase of
a SIB/SIF.  It is important to note that our emphasis on $g({\phi})$
does not ignore $g(r)$.  Rather it gives equal importance of $g(r)$ and
$g(k)$ (the interparticle momentum correlation) because of $\phi =
\textbf{k}.\textbf{r} = 2n\pi$, while the emphasis on only $g(r)$ ignores $g(k)$. 
\par
\noindent
\textit{4.5. $<\Phi_n|V_{HC}(r)|\Phi_n>$ } :
As reported in \cite{4,5}, one obtains a non-zero $<\Phi_n|V_{HC}(r)
\equiv A.\delta (r)|\Phi_n>$ when he uses
\begin{equation}
\Phi_n = {1\over\sqrt{N!}}\sum_{P}(\pm 1)^{[P]}
[u_{P\alpha_i}(r_1)u_{P\alpha_j}(r_2)...
u_{P\alpha_N}(r_N)]   \label{eq17}
\end{equation}
\noindent
to represent a state of $N$ particles and evaluates the integral
over $r_1$ and $r_2$.  However, in this process he does not
encounter $\psi_k{(r)}^-$ to implement $\psi_k{(r)}^-|_{r=0} = 0$, -the
basic character of a pair waveform.  Naturally, this process does not
ensure the accuracy of non-zero $<\Phi_n|V_{HC}(r)|\Phi_n>$ so
obtained \cite{4,5}.  Guided by this observation, we rearrange Eqn.(17) as
\begin{eqnarray}
\Phi_n &=& {1\over\sqrt{N!}}\sum_{i<j}
\left\{u_{\alpha_i}(r_1)u_{\alpha_j}(r_2)
\pm u_{\alpha_j}(r_1)u_{\alpha_i}(r_2)\right\}. \nonumber\\
&&\sum_{P'}(\pm 1)^{[P']} [u_{P'\alpha_l}(r_3) ... u_{P'\alpha_N}(r_N)]
\label{eq18}
\end{eqnarray}
\noindent
which resolves $\psi_k{(r)}^-$ and $\exp{(i\textbf{ K}.\textbf{ R})}$
waveforms of two particles at $r_1$ and $r_2$ (say) and evaluate
the $<\Phi_n|V_{HC}(r)|\Phi_n>$ related integral over $r$ and $R$.  
Note that $P'$ in Eqn.(18) represents the permutation over different
$\alpha$ includes all $N$ $\alpha$ (\textit{ i.e.} $\alpha_1$, $\alpha_2$,
...$\alpha_N$) except those identified as $\alpha_i$ and
$\alpha_j$ (with integers $i$ and $j$ running from 1 to $N$).  Defining  
\begin{eqnarray}
\psi_{\alpha_i, \alpha_j}(r,R)^{\pm} &=& [u_{\alpha_i}(r_1)u_{\alpha_j}(r_2)
\pm u_{\alpha_j}(r_1)u_{\alpha_i}(r_2)]  \nonumber\\
&=& \psi_{\alpha_i, \alpha_j}(r)^{\pm}
\exp{[i(\textbf{k}_{i}+\textbf{k}_{j}).\textbf{R}]}   \label{eq19}
\end{eqnarray}
\noindent
we find that  
\begin{equation}
<\Phi_n|\delta{(r)}|\Phi_n> = {N(N-1)\over{2}}
I_{ij, i'j'}^{(1)}.I_{ij, i'j'}^{(2)}.I^{(3)} \label{eq20}
\end{equation}
\noindent
with  
\begin{eqnarray}
I_{ij, i'j'}^{(1)} &=& \int \psi_{\alpha_i, \alpha_j}(r)^c\delta{(r)}
\psi_{\alpha_i', \alpha_j'}(r)^c d^3r\nonumber\\
&=& \psi_{\alpha_i, \alpha_j}(r)^c|_{r=0}
.\psi_{\alpha_i', \alpha_j'}(r)^c|_{r=0} = 0.       \label{eq21}
\end{eqnarray}
\noindent
Here $\psi_{\alpha_i, \alpha_j}(r)^c$ is that $\psi_{\alpha_i,
\alpha_j}(r)^{\pm}$ which vanishes at $r=0$ since only such a waveform
can correctly represent two HC particles (bosons/fermions).  We also have
\begin{eqnarray}
I_{ij, i'j'}^{(2)}&=&
\int \exp{[-i(\textbf{k}_i+\textbf{k}_j).\textbf{R}]} \nonumber\\
&&.\exp{[i(\textbf{k}_{i'}+\textbf{k}_{j'}).\textbf{R}]}d^3R \nonumber\\
&=& \delta_{(\textbf{k}_i+\textbf{k}_j),(\textbf{k}_{i'}+
 \textbf{k}_{j'})}   \label{eq22}
 \end{eqnarray}
\noindent
representing the conservation of momentum of two particles in their
collision, and
\begin{eqnarray}
I^{(3)} = <\Phi_n'|\Phi_n'>
&=& {1\over{(N-2)!}}\sum_{P'}\sum_{Q'}(\pm 1)^{[P']+[Q']} \nonumber\\
&&[\delta_{P'\alpha_3, Q'\alpha_3} ...
\delta_{P'\alpha_N, Q'\alpha_N}]   \label{eq23}
\end{eqnarray}
\noindent
where $\Phi_n'$, -a part of Eqn.(18), can be expressed as
\begin{eqnarray}
\Phi_n'(r_3, ..., r_N) &=& {1\over\sqrt{(N-2)!}}\sum_{P'}(\pm 1)^{[P']}
[u_{P'\alpha_3}(r_3) \nonumber\\
&&u_{P'\alpha_4}(r_4)... u_{P'\alpha_N}(r_N)]. \label{eq24}
\end{eqnarray}
\noindent
Evidently, $<\Phi_n|A.\delta{(r)}|\Phi_n>$ vanishing 
due to zero value of $I_{ij, i'j'}^{(1)}$ (Eqn.21) renders
\begin{equation}
<\Phi_n|A.\sum_{a<b}\delta{(r_b - r_a)}|\Phi_n> =  0  \label{eq25}
\end{equation}
\noindent
which differs from its non-zero value \cite{4,5}.  It is important to note
that the zero value of $<\Phi_n|A.\delta{(r)}|\Phi_n>$ should not
be confused to imply that energy eigenvalues of a system of non-interacting
particles are identical to those of a system of particles interacting
through $V_{HC}(r)$.  While particles in the former case have no
way to identify the presence of each other and the lowest possible $q =
q_o = \pi/L$ and corresponding energy $\varepsilon_o = h^2/8mL^2$ are
decided by the size $L$ of the container, $q_o$ and $\varepsilon_o$, in the
latter case (\textit{cf.} Section 4.2), are decided by $d$ which is not only
much shorter than $L$ but is also decided by net sum of $V(r_{ij})$.
\par
\noindent
\textit{4.6. Energy eigenvalue} : It is evident that a particle in a
SIB/SIF can be represented more accurately by a
$\Psi{(r, R)}^-$ pair waveform (Eqn.13) than a plane wave due to its 
wave mechanical superposition with itself (self superposition)
or with another particle (mutual superposition) in a process of
its collision; self superposition of a particle can also be visualized
when it returns after its collision with the walls of the container.
However, the consistency of this representation demands that the kinetic
energy terms of $H^o(N)$ (Eqn.2) should also be paired as
\begin{equation}
h(i) = {1\over{2}}\left[h_i + h_{i+1}\right] \label{eq26}
\end{equation}
\noindent
with $h_{N+1} = h_1$ and     
\begin{equation}
h_i = -{\hbar^2\over{2m}}{\bigtriangledown}^2_i \quad {\rm and}
\quad h(i) =  -{\hbar^2\over{8m}}{\bigtriangledown}^2_{R_i}
-{\hbar^2\over{2m}}{\bigtriangledown}^2_{r_i} \label{eq27}
\end{equation}
\noindent
Naturally, this renders
\begin{equation}
H^o(N) = \sum_i^N h(i) + \sum_{i<j}^N A.\delta{(r_{ij})}. \label{eq28}
\end{equation}
\noindent
One can also use other schemes of pairing $h_i$ terms of
$H^o(N)$ as discussed in \cite{15}.  The rearrangement of $H^o(N)$ as
expressed by Eqn.(28) and representation of each particle by a
$\Psi{(r, R)}^-$ type pair waveform facilitate in constructing a
wave function for a state of $N$ particle system by following the
standard procedure.  This is particularly so because expectation
value of $V_{HC}(r_{ij}) \equiv A.\delta{(r_{ij})}$ vanishes (\textit{ cf.}
Eqns. 16 and 25).  For $N$ particles we have $N$ different
$\Psi{(r, R)}^-$ rendering $\Sigma = N!$ different $\Psi_n$ 
(through permutations of $q_i$ or $r_i$ of N particles) for $n-$th
state of equal energy $E_n = E_n(K) + E_n(k)$.  Defining
\begin{equation}
E_n(K) = \sum_i^N \varepsilon (K)_i \quad {\rm and}
\quad E_n(k) = \sum_i^N \varepsilon (k)_i  \label{eq29}
\end{equation}
\noindent
we have 
\begin{equation}
\Psi_n =  \phi_n(q).\phi_n(K)  \label{eq30}
\end{equation}
\noindent
with 
\begin{equation}
\phi_n(q) = \left[ \left({2\over{\rm V}}\right) ^{N\over{2}}
\prod_{i=1}^{N} \sin{(\textbf{q}_i.\textbf{r}_i)} \right]
\exp[-iE_n(k)t/{\hbar}]  \label{eq31}
\end{equation}
\noindent
and 
\begin{eqnarray}
\phi_n(K) &=& {A}.{\left( {1\over{\rm V}}\right)}^{N\over{2}}
{\sum_{\textit{ p}K}}{(\pm 1)^p} \prod_{i=1}^{N}
\exp [i(\textbf{K}_i.\textbf{R}_i]. \nonumber\\
&&\exp[-iE_n(K)t/{\hbar}]   \label{eq32}
\end{eqnarray}
\noindent
with $A={\sqrt {1/N!}}$.  Here $\sum_{\textit{ p}K}{(\pm 1)^p}$ refers to the
sum of different permutations of $K$ over all particles.  While the use of
$(+1)^p$ or $(-1)^p$ in Eqn.(32) depends on the bosonic or fermionic
nature of the system for their spin character of the particles, the use of
the restriction $q_i \ge \pi/d$ in Eqn.(31) treats the so called fermionic
behavior (in the $r-$space) of HC particles (bosons and fermions
alike).  Evidently, a state function of $N$ HC bosons should differ from
that of $N$ HC fermions in the choice of $(+1)^p$ or $(-1)^p$.  Note that
$\Sigma$ different $\Psi_n$ counted above take care of the permutation of
$k=2q$.  We have
\begin{equation}
\Phi_n ={1\over{\sqrt{\Sigma}}}.\sum_i^{\Sigma} \Psi_n^{(i)} \label{eq33}
\end{equation}
\noindent
which represents the general form of a state function that should reveal 
the physics of a SIB/SIF.  Note that ${\Phi_n}$ represents a state where
each particle, as a WP of size $=\pi/q$, has a plane wave motion of
momentum \textbf{K}.  Since $\phi_n(q)$ (Eqn.31) appearing in each
$\Psi_n^{(i)}$ of Eqn.(33) through Eqn.(30) vanishes at every point 
$r_i = r_j$, that defines the CM of $i-$th and $j-$th particles, one
may easily find 
\begin{equation}
<\Phi_n|V_{HC}(r_{ij}) \equiv A.\delta (r_{ij})|\Phi_n> = 0 \label{eq34}
\end{equation}
\noindent
and  
\begin{eqnarray}
<\Phi_n|H^o(N)|\Phi_n> &=&
\sum_i^N {\hbar^2\over{8m}}\left(K_i^2 + k_i^2\right)  \nonumber\\
&=& \sum_i^N {\hbar^2\over{8m}}\left(K_i^2 + 4q_i^2\right)
\label{eq35}
\end{eqnarray}
\noindent
which can be shown \cite{15} to render
\begin{equation}
E_o = Nh^2/8md^2 = N{\varepsilon}_o    \label{eq36}
\end{equation}
\noindent
as the G-state energy for a SIB for which all $K$ values can be zero.
However the G-state energy for a SIF, where only two particles (with spin
up and spin down) can occupy a state of an allowed $K$, is found to be
\cite{16,17}
\begin{equation}
E_o =  N{\varepsilon}_o  + E_F  \label{eq37}
\end{equation}
\noindent
with $E_F =  h^2/8\pi{m}(N/1.5045{\rm V})^{2/3}$ representing the Fermi
energy of $N$ particles derived for their $K$ motions.  Note that $K$
motion of a particle in SMW configuration is a kind of free
particle motion with effective mass being $4m$.
\par
\noindent
\textit{4.7. $V^{\rm A}(r_{ij})$ as perturbation} : It is obvious that
$V^{\rm A}(r_{ij})$ can affect the relative configuration of particles
in a SIB/SIF which, in the G-state of the system, is defined by the
nearest neighbor separation $d$ and the least possible $q$ of a
particle $q_o = \pi/d$.  Evidently,
when this state of the system is subjected to a perturbation by
attractive potential, $V^{\rm A}(r_{ij})$, its G-state energy
$E_o = N{\varepsilon}_o = Nh^2/8md^2$ changes to $E'_o = N{\varepsilon}'_o
= Nh^2/8md'^2$ rendering a net fall in G-state energy by 
\begin{equation}
E_g ={Nh^2\over{8m}}\left[{1\over{d^2}} - {1\over{d'^2}}\right]
       \approx {Nh^2\over{4m}}{{d'-d}\over{d^3}}  \label{eq38}
\end{equation}
\noindent
where $d'$ represents the increased $d$ value which arises due to
zero point repulsion coming into effect at a temperature
at which particles in the system satisfy $\lambda_T = 2d$ (with
thermal de Broglie wave length, $\lambda_T = h/\sqrt{2\pi mk_BT}$).  In
fact this force leads the system to have volume expansion on its cooling
below certain temperature as observed experimentally for LHe-4 as well
as LHe-3.  We define $E_g$ as the
energy gap between the normal and superfluid states of the system.  It
also represents a kind of collective binding of all atoms for which the
entire system can be identified as a single macro-molecule.  
\par
\noindent
\textit{4.8. Effective interparticle interaction} :
It may be noted that our approach of developing the microscopic theory
not only replaces $V^{\rm R}(r_{ij})$, as an approximation, by
$V_{HC}(r_{ij}) \equiv A.\delta{(r_{ij})}$ but also imposes a condition,
``that two WPs of HC particles
should not share any point \textbf{ r} in configuration space'', -equivalent
to assuming the presence of a repulsion of \textit{ finite range}, $r_a =
\lambda/2$; it appears that the WP manifestation of particle extends the 
range of the influence of $V_{HC}(r)$ from $r_a = \sigma$ to $r_a =
\lambda/2$ when $\lambda/2 > \sigma$.  This repulsion is nothing but
the zero point repulsion \cite{29} which can be derived as the first $d$
derivative of $\varepsilon_o = h^2/8md^2$ representing the G-state energy of
a particle. This again shows that $A$ in $V_{HC}(r) \equiv A.\delta{(r)}$
should be $h^2/8md^2$.
\par
Further, since $V^{\rm R}(r_{ij})$ in most SIB  falls faster (in LHe-4
it varies as $r^{-12}$) than the zero-point repulsion, varying as $r^{-2}$,
the latter would dominate $V^{\rm R}(r_{ij})$ particularly for all $r >
\sigma$ and $\le \lambda/2$ and this observation agrees with the
experimental facts that : (i) LHe-4 and LHe-3 do not solidify due to
zero-point repulsion even at $T=0$ unless they are subjected to an external
pressure of $\approx 25$ and $\approx 30$ atms, respectively, and
(ii) they exhibit volume expansion with falling $T$ around 2.2 and 0.5 K,
respectively\cite{30}.  It may also be noted that our condition,
$\lambda/2 \le d$, identifies $d$ as the upper limit of the WP size
$\lambda/2$ (the key aspect of our theory) of a particle and $d$ is
decided by the net sum of $V(r_{ij})$ without any approximation.  Evidently,
these observations prove that \textit{ our theory accounts for the
$V^{\rm R}(r_{ij})$ and $V^{\rm A}(r_{ij})$ components of
$V(r_{ij})$ close to their real effect.}
\par
\noindent
\textit{4.9.  Why (\textbf{q}, -\textbf{q}) pair condensation }: 
The phenomenon of superfluidity/superconductivity of a fermionic
system is attributed to the condensation of Cooper pairs of fermions
for a reason that the Pauli exclusion
principle forbids two identical fermions from occupying single energy state,
while any number of these pairs presumed to behave like bosons can do so.
Because Pauli exclusion does not apply to bosons, conventional theorists  
find no difficulty in assuming the condensation of macroscopically
large number of bosons into a single particle state of $p = 0$ as their
main theme.  However, this assumption ignores the fact that the way 
two fermions do not occupy same point in $k-$space, two HC particles do
not occupy same point in $r-$space.  This is particularly important
because the requirement of antisymmetry of two fermion wave function,
$\Phi_a{(1,2)} = [v_{\textbf{k}'}(\textbf{r}_1).v_{\textbf{k}''}
(\textbf{r}_2) - v_{\textbf{k}'}(\textbf{r}_2).v_{\textbf{k}''}
(\textbf{r}_1)]$, for their exchange, makes $\Phi_a{(1,2)}$ vanish not only
for $\textbf{k}' = \textbf{k}''$ but also for $\textbf{r}_1 =
\textbf{r}_2$.  Evidently, if $\Phi{(1,2)}$ of two HC particles is subjected
to a condition that it should vanish for $\textbf{r}_1 = \textbf{r}_2$, 
$\Phi{(1,2)}$ has to be identically antisymmetric and would, obviously,
vanish also for $\textbf{k}' = \textbf{k}''$.  This implies that
two HC quantum particles in $r-$space behave like two fermions behave in
$k-$space and concludes that two HC particles (\textit{excluded to have}
$\textbf{r}_1 = \textbf{r}_2$) can not have $\textbf{k}' = \textbf{k}''$, particularly,
in a state of their \textit{wave mechanical superposition} (\textit{i.e.} a quantum
state of $\lambda > d$).  Note that the inference would be valid not
only for particles of $\sigma \approx 0$ (\textit{i.e.} particles interacting
through $\delta$-function repulsion) but also for $^4He$ type atoms because
finite size HC repulsion becomes equivalent to $\delta$-function repulsion
for particles of $\lambda/2 > \sigma$ \cite{14}.  However, we also note that
there is a difference in Fermi behavior due to HC nature and that due to
half integer spin; while the former excludes every particle from
having $q < \pi/d$ (applies identically to HC bosons and HC fermions), the
latter excludes two particles (applies to fermions only) from having equal
\textbf{K}.  Evidently, this excludes the possibility of  
nonzero $n_{p = 0}(T)$ which has also been shown to be \textit{inconsistent with
excluded volume condition of} HC \textit{particles} \cite{28}.
\par
\textit{As such like Pauli exclusion provides effective repulsion to keep two
fermions apart} \cite{20}, the \textit{volume exclusion condition applicable to} HC \textit{quantum particles} and 
WP \textit{manifestation of quantum particles render such repulsion to keep
their} WPs \textit{ at $r \ge \lambda/2$}; experimentally observed volume
expansion of LHe-4 with decreasing $T$ near $T_{\lambda}$ \cite{30}
corroborates this fact.
Evidently, there is no doubt that superfluidity of LHe-4 type SIB
originates from the condensation of (\textbf{q}, -\textbf{q}) pairs.  The
\textit{binding between two particles originates from their inherent interatomic
attraction} and this has been discussed in detail in Section (5.4) of
\cite{15}.
\section{5.0.  Conclusion}
As such, this study concludes that : (i) Under the approximation 
$V_{HC}(r) \equiv A.\delta{(r)}$, $\Psi{(r_1, r_2)}$ (Eqn. 11) (for
the situation $r > \lambda/2$) and $\psi_k{(r,R)}^-$ (Eqn. 13) (for
$r \approx \lambda/2$) represent exact solutions of Eqns.(3) and (4),
respectively, (ii) allowed $q$ values satisfy $q \ge \pi/d$
(\textit{cf.} Section 4.3), (iii) $\phi$-correlations dominates the
behavior of low temperature phase (\textit{cf.} Section 4.4), and
(iv) $<\Phi_n|A.\delta{(r)}|\Phi_n>$ has zero value (\textit{cf.}
Section 4.5) which differs significantly from its non-zero value
concluded through conventional formulations\cite{4,5}.  The errors
of non-zero value of $<\Phi_n|A.\delta{(r)}|\Phi_n>$ are also evident from
its dependence on momentum and spin distribution of particles\cite{4,5} 
which does not agree with the fact that $A.\delta{(r = 0)} = \infty$ 
and $A.\delta{(r > 0)}
= 0$ is independent of momentum and/or spin states of two HC particles.
As such the main sources of our basic problems in developing a correct
theory of a LHe-4/LHe-3 have been the finer intricacies of wave mechanics
(\textit{cf.} Sections. 4.2-4.6 and Eqns.(16 AND 25) that we, somehow, 
missed for so long.  This author
too never imagined of this fact.  Only recently, we identifies
these intricacies and developed a new approach that resolves
these problems and helps in finding an almost exact theory of a
LHe-4(LHe-3) type SIB(SIF)\cite{15,16,17}.  As such the present analysis
provides strong foundations to our approach. 
We also find that : (i) our approach not only uses $V^{\rm R}(r_{ij})$
and $V^{\rm A}(r_{ij})$ components of $V(r_{ij})$ effectively but also
analyzes the way zero-point repulsion dominates the behavior of a 
superfluid SIB (\textit{cf.} Section (4.8)) and (ii) there are compelling
reasons (\textit{cf.} Section (4.9)) for a SIB too to have (\textbf{q},
-\textbf{q}) pair condensation.  
\begin{acknowledgements}
The author is thankful to Prof. M.K. Parida for
useful discussion.
\end{acknowledgements}


\begin{thebibliography}{99}
\bibitem{27}
For a pair described by $\psi_k{(r)}^-$ [\textit{cf.} Eqn.(15)] (with
$k = 2q$), it can be shown that $<{\bigtriangledown}^2_r> \equiv <q^2> = q^2$
and $<{\bigtriangledown}_r> \equiv <q> = 0$ rendering $\Delta{k} =
\sqrt{<k^2> - <k>^2}  = k$.  One can also obtain  $<r> = <k,r|r||k,r>/
<k,r|k,r> = \lambda/2$  by using $|k, r> \equiv \psi_k(r)^-$ (Eqn.15)
and the fact that $r$ can be taken to vary from $r = 0$ (when P1
and P2 collide at their CM) to $r = \lambda$ (when P1 and P2 are $\lambda/2$
distance away from their CM) as the limits of $r$ involved in the
integration.  Here we presume that $\psi_k(r)^-$ is a s-wave state
(\textit{i.e.}, \textbf{k}.\textbf{r} = $kr$) of P1 and P2 and these particles always
stay within two neighboring anti-nodal regions of $\psi_k(r)^-$ as happens
in superfluid $^4He$ \cite{15}.  However, if particles can move out of
$r = \lambda$ limit, we can only conclude that $<r> \ge \lambda/2$ which
applies to normal liquid $^4He$ \cite{15}.
\end{thebibliography}
\end{document}